# 4π detector for study of Zeno effect using $^{220}$Rn → $^{216}$Po α-α correlated chains

*L.Nadderd*[*]*, Yu.S.Tsyganov, K.Subotic*[*]*, A.N.Polyakov, Yu.V.Lobanov, A.V.Rykhlyuk*

**Abstract**

First test of the 4π detector for study of exponential law of radioactive decay and possibility of observation of Zeno effect [1-3], measuring the mean life of $^{216}$Po is presented. This detector consists of two surface-barrier n-Si(Au) detectors placed in the close contact ( <1mm) to each other. Th foil is used as a generator of Rn- gas. The foil was covered by nuclear filter with 0.4 μm pores and placed in the bottom of vacuum chamber. So, direct alpha-particles can not reach the detector active surface. The measured half-live of $^{216}$Po of 143.5 (0.6) ms is in agreement with literature sources. It is shown, that for the definite source the random background correlation counts are mostly caused by the $^{220}$Rn - $^{216}$Po alpha activities and that it is not an easy task to decrease a background level of units of percents in the time region $>4T_{1/2}$. Both, the data acquisition system and the vacuum chamber design are presented in brief. Authors plan to perform the second set of tests at the end of 2007 and beginning of 2008 years.

1. **Introduction**

The consideration of general problems of decay of the unstable systems by Khalfin [1] demonstrated that decay rate of a quasi-stationary state does not obey exactly the exponential law. He demonstrated that the decay rate was slowing-down for long times. Fonda et al. [2] have analyzed a decay model and showed that in some circumstances significant deviation from the exponential law could occur at about 10 lifetimes and inverse power-law domination could occur at about 25 lifetimes.

2. **1$^{st}$ step in the Rn → Po α-α correlation test measurements: effect and backgrounds.**

To measure correlations the regions $t > T_{1/2}*n1$, and $t < T_{1/2}/n2$, ( $n_{1,2} >>1$) the setup shown schematically in the fig.1 has been designed. Both silicon detectors are made from n-silicon of about 1.2 KOhm·cm receptivity wafers with evaporated Au top electrodes. As to rear contacts, they are produced by chemical extraction of Ni onto non-etched silicon surface. Leakage current at the working bias of 20 Volts was about 0.4μA. The geometrical detection efficiency was estimated to be 80% from 4π. We use Th source when no direct alpha particle could be detected by both detectors, but only after gaseous transport of Rn single atoms onto the sensitive detector area.

---

[*] Institute **VINCA**, Belgrade, Rep. Serbia

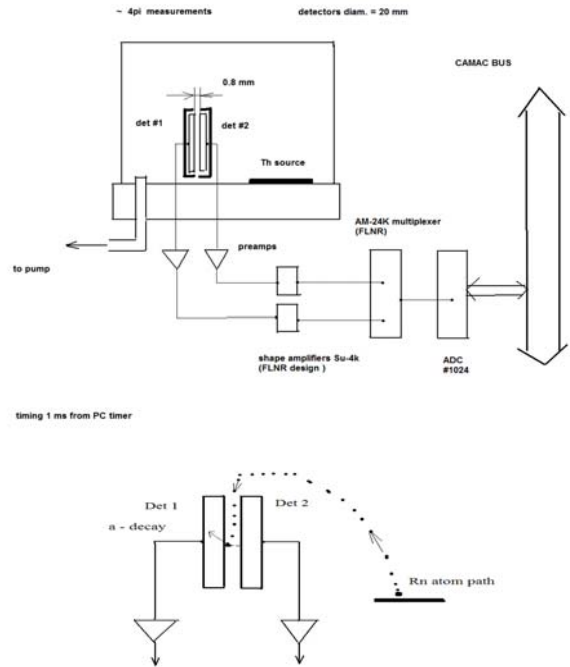

**Fig.1** Schematic view of the 4π detector. No direct alpha-particles could reach to the active surfaces of both detectors (picture down: way of an Rn atom to detector active Au-surfase ).

## 3. Measured spectra for $^{220}$Rn $\rightarrow$ $^{216}$Po α-decays

In the fig.2 the measured spectra for both detectors are shown. Note, that he peaks under interest are located in middle, whereas left and right peaks are of 6.06 and 8.78 MeV which are used for calibration process. In the fig.3 two – dimensional alpha-alpha correlation spectrum for 650 ms time interval is shown. They are definitely shown an existence of backgrounds of different nature: random correlation for Rn$\rightarrow$Po links as well as non-related counts. The acquisition time for the given spectra was about one day.

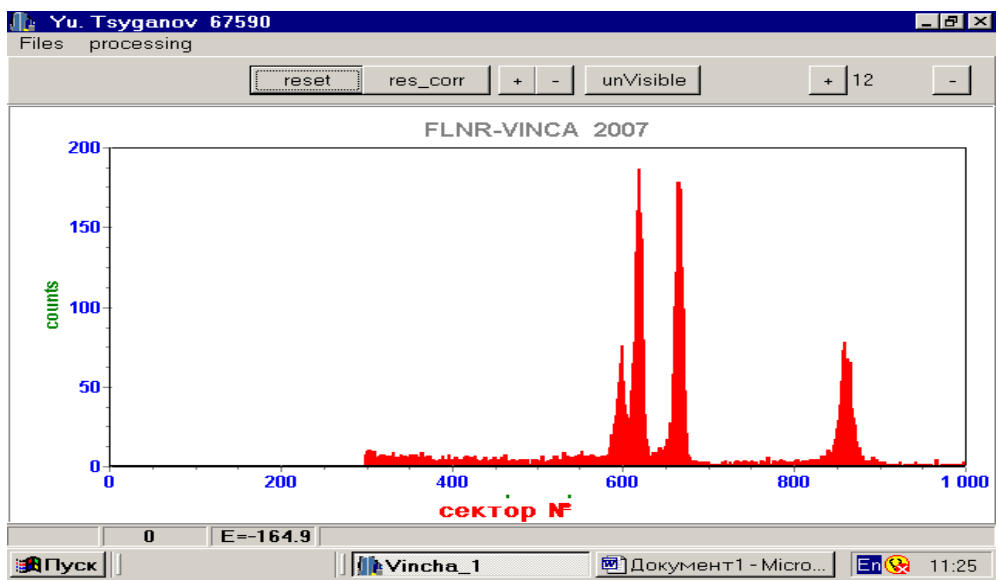

**Fig.2** Spectrum for $^{220}$Rn$\rightarrow$$^{216}$Po correlated chains. Two intense peaks – Rn && Po. Two others 6.062 and 8.785 MeV

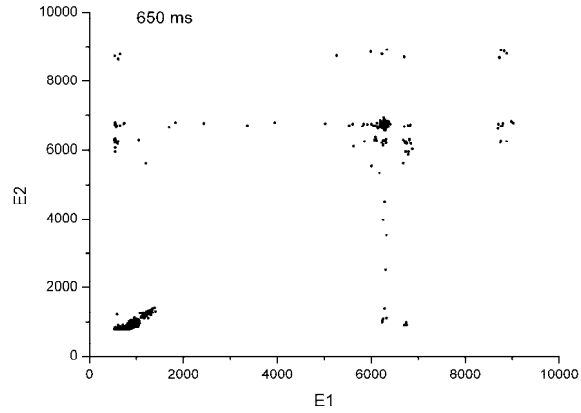

**Fig.3** *E1-E2* correlation spectrum. Time window is 650 ms. Both scales(y,x) in KeV's. Group of events symmetrical according to y=x line is in fact background is explained for Rn→Po decays themselves.

## 4. Discussion

Although units of percents of background explained itself random correlation Rn→Po are existed even in the range more than four half-lives, such the effect can be described in the form of time independent constant, due to the rate of both sources (Rn & Po) are constant in the mean sense. So, the mathematical description of background right tail is a great of interest. Of course, the additional sources which are not related to the $^{220}$Rn→$^{216}$Po random links are also should be under consideration. From the other hands, any attempts to minimize the mentioned factor are required a lot of exposition time, like months or even years.

## 5. Conclusions

First test of simple 4π- detector to study Zeno effect for Rn→Po correlation sequences has been performed. It is shown for times > $4T_{1/2}$ background of random correlation is not less that of units of percents. There are three reasonable ways for prolongation of the experiment:

   a) by using source with much more less activity → very long time of measurement will be required;
   b) by using the same or even much more powerful source, but with exact mathematical description of background shape in the t > 10 $T_{1/2}$ region;
   c) by using special design detector pair with position sensitive resistive layers, but with the slit as small as possible, to prevent Po atom to go out the region restricted by a several mm units;

**Appendix1**

When preparing this manuscript, measurements with larger statistics (16 days) were performed[*]. In the Fig. 4 decay curve for $^{220}$Rn-$^{216}$Po α-α correlation chains is shown. Measured parameters of half-life and random background are shown in the frame.

---

[*] The DGFRS spectrometer (two spectrometry channels of "veto" detector) is used

An accuracy of an elapsed time measurement was of one microsecond. Both time decay constants are correspond to each other (k- parameter in the exponential function fit ).

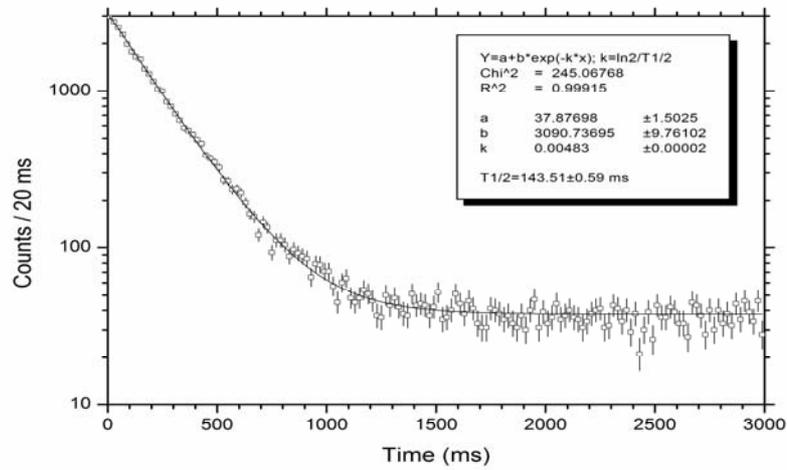

a)

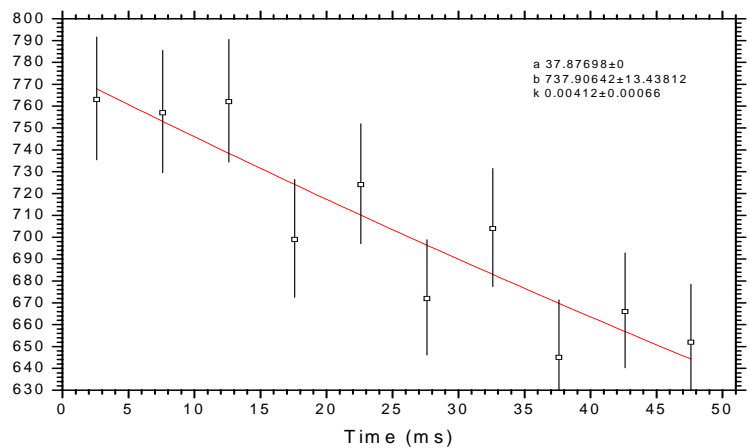

b)

**Fig.4** Decay curve for $^{220}$Rn → $^{216}$Po α-α correlated sequences **a)**.
Measured half-life value is equal to 143.51±0.59 ms
(The same value reported in [4] is equal to 145 ±*2* ms.)
**b)** the same exponential decay, but for (0, 50 ms) time interval is taken into account